\documentclass[3p,preprint,hyperref]{elsarticle}
\input{prolegomenaHP}

\begin{document}

\begin{frontmatter}
\title{Elliptic Polylogarithms and Basic Hypergeometric Functions}

\author[torino]{Giampiero Passarino\fnref{support}}
\ead{giampiero@to.infn.it}

\address[torino]{\csumb}

\fntext[support]{\support}

\begin{abstract}
Multiple elliptic polylogarithms can be written as (multiple) integrals of products of
basic hypergeometric functions. The latter are computable, to arbitrary precision, using a 
$q\,$-difference equation and $q\,$-contiguous relations.
\end{abstract}

\end{frontmatter}

\section{Introduction}
There is a wide class of Feynman integrals, mainly related to massless theories, which can 
be expressed in terms of multiple polylogarithms. More challenging are Feynman integrals, which
cannot be expressed in terms of multiple polylogarithms. Evaluating this integrals one encounters 
elliptic generalizations of (multiple) polylogarithms (EP); examples can be found in
\Brefs{Adams:2015ydq,Adams:2016sob,Adams:2016xah} and in 
\Brefs{Bloch:2014qca,Bloch:2016izu}. From a more abstract point of view on (multiple)
elliptic polylogarithms, in particular on their analytic structure, see \Brefs{Lev,BL}
and \Brefs{Beil,Bloch,Zag,Bloch:2013tra}.
An interesting problem is to find a suitable integral representation and the analytic 
continuation of EPs and an efficient algorithm for their numerical evaluation.
\section{Elliptic polylogarithms}
In \Brefs{Adams:2015ydq,Adams:2016sob,Adams:2016xah} the following functions, of depth
two, are defined:
\bq
\ELi{n}{m}{x}{y}{q} = \sum_{j=1}^{\infty}\,\sum_{k=1}^{\infty}\,\frac{x^j\,y^k}{j^n\,k^m}\,
q^{jk} \spp
\label{elione}
\eq
For the sake of simplicity we shall assume that $\mathrm{arg}(q) = 0$. The results derived can 
be extended by using analytic continuation. Following \Bref{BL} we prefer to start from
\bq
\ELi{n}{m}{x}{y}{q} = \sum_{j=1}^{\infty}\,\frac{x^j}{j^n}\,\li{m}{y\,q^j} \spc
\label{elitwo}
\eq
where $\li{m}{\mrz}$ is a generalized polylogarithm~\cite{Kolbig:1983qt}. 
Furthermore, $q \in \Cf^{x}$ with $| q | < 1$ and $y \in \Cf$ with $1 \notin q^{\Rf}\,y$. 
Thus, for $x < | q |^{-1}$ the series converges absolutely. It is immediately seen that 
$\ELi{0}{0}{x}{y}{q}$ has a pole at $x = 1$. 

Note that \Bref{BL} defines functions whose simplest example is given by
\bq
\mrE(x\,,\,y\,;\,q) = \sum_{j \in \Zf}\,x^j\,\li{1}{y\,q^j} \spc
\eq
requiring $1 < x < | q |^{-1}$. The elliptic dilogarithm is defined in \Bref{Bloch:2016izu} as
\bq
{\hat{\mrE}}_2(x) = \sum_{n=0}^{\infty}\,\Bigl[ \li{2}{q^n\,x} - \li{2}{ - q^n\,x} \Bigr] -
\sum_{n=1}^{\infty}\,\Bigl[ \li{2}{\frac{q^n}{x}} - \li{2}{- \frac{q^n}{x}} \Bigr] \spc
\eq
and we have the following relations:
\bqa
\sum_{n=1}^{\infty}\,\li{2}{q^n\,\mrz} &=& \ELi{0}{2}{1}{\mrz}{q} \spc
\nl
\Lambda_2( - i\,\mrz ) &=& \frac{1}{2\,i}\,\Bigl[ \li{2}{\mrz} - \li{2}{- \mrz} \Bigr]= 
\sum_{n=0}^{\infty}\,\frac{(- 1 )^n}{( 2\,n + 1 )^2}\,( - i\,\mrz )^{2\,n + 1} \spc
\eqa
where the last relation holds for $| \mrz | < 1$. 

Next, generalizations of higher depth of \eqn{elitwo} are also defined in \Bref{Adams:2016xah}. 
They are defined by
\bqa
\mathrm{ELi}_{[n]_l\,;\,[m]_l\,;\,[2\,\sigma]_{l-1}}
\lpar [x]_l\,;\,[y]_l\,;\,q \rpar &=&
\sum_{j_1=1}^{\infty}\,\dots\,\sum_{j_l=1}^{\infty}\,
\sum_{k_1=1}^{\infty}\,\dots\,\sum_{k_l=1}^{\infty}\,
\frac{x^{j_1}}{j^{n_1}_1}\,\dots\,\frac{x^{j_l}}{j^{n_l}_l}\;
\frac{y^{k_1}}{k^{m_1}_1}\,\dots\,\frac{y^{k_l}}{k^{m_l}_l}
\nl
{}&\times&
\frac{q^{j_1\,k_1 +\,\dots\,+ j_l\,k_l}}
{\prod_{i=1}^{l-1}\,\lpar j_i\,k_i +\,\dots\, + j_l\,k_l \rpar^{\sigma_i}} \spc
\eqa
which are elliptic polylogarithms of depth $2\,l$. We have introduced the abbreviation
\bq
[n]_l = n_1\,,\,\dots\,,\,n_l 
\eq
\etc The following relation holds:
\bq
\ELi{n}{m}{x}{y}{q} = \int \frac{d\mrz}{\mrz}\,\ELi{n-1}{m}{\mrz\,x}{y}{q} = 
\int \frac{d\mrz}{\mrz}\,\ELi{n}{m-1}{x}{\mrz\,y}{q} \spp
\label{reda}
\eq
Furthermore, for $l > 1$, we can use
\bqa
\int_0^1 \frac{d\mrz}{\mrz}\,
\mathrm{ELi}_{[n]_l\,;\,[m]_l\,;\,[2\,\sigma]_{l-1}} 
\lpar [x]_l\,;\,[y]_l\,;\,\mrz\,q \rpar &=&
\mathrm{ELi}_{[n]_l\,;\,[m]_l\,;\,2\,(\sigma_1 + 1)\,\dots\,2\,\sigma_{l-1}} 
\lpar [x]_l\,;\,[y]_l\,;\,q \rpar \spc
\nl
\label{redb}
\eqa
\bqa
\mathrm{ELi}_{[n]_l\,;\,[m]_l\,;\,0\,\dots\,2\,\sigma_{l-1}} 
\lpar [x]_l\,;\,[y]_l\,;\,q \rpar &=&
\ELi{n_1}{m_1}{x_1}{y_1}{q}\,
\mathrm{ELi}_{n_2\,\dots\,n_l\,;\,m_2\,\dots\,m_l\,;\,2\,\sigma_2\,\dots\,2\,\sigma_{l-1}} 
\lpar x_2\,\dots\,x_l\,;\,y_2\,\dots\,y_l\,;\,q \rpar \spp
\label{redc}
\nl
\eqa
In \eqns{reda}{redc} we find recurrence relations where the starting point is always
$\ELi{0}{0}{x}{y}{q}$, \eg
\bq
\mathrm{ELi}_{n_1\,,\,n_2\,;\,m_1\,,\,m_2\,;\,2} 
\lpar x_1\,,\,x_2\,;\,y_1\,,\,y_2\,;\,q \rpar =
\int_0^1 \frac{d\mrz}{\mrz}\,
\ELi{n_1}{m_1}{x_1}{y_1}{\mrz\,q}\,
\ELi{n_2}{m_2}{x_2}{y_2}{\mrz\,q} \spc
\eq
followed by repeated applications of \eqn{reda}. In the next Section we will define
basic hypergeometric series and establish the connection with elliptic polylogarithms.
\section{Basic hypergeometric functions}
It is easy to show that $\ELi{0}{0}{x}{y}{q}$ can be written in terms of a basic hypergeometric 
function. Indeed,
\bqa
\ELi{0}{0}{x}{y}{q} &=& 
   \sum_{j=1}^{\infty}\,x^j\,\frac{y\,q^j}{1 - y\,q^j} =
   \frac{ x y q}{1 - y\,q}\,{}_2\,\upphi_{1}\lpar q\,,\,y q\,;\,y q^2\,;\,q\,,\,x q \rpar 
\nl
{}&=&
   \sum_{j=1}^{\infty}\,y^j\,\frac{x\,q^j}{1 - x\,q^j} =
   \frac{ x y q}{1 - x\,q}\,{}_2\,\upphi_{1}\lpar q\,,\,x q\,;\,x q^2\,;\,q\,,\,y q \rpar \spc
\label{eli00def}
\eqa
where the basic hypergeometric series is defined as
\[
{}_{r}\upphi_{s} \left[
\begin{array}{c}
\alpha_1,\,\dots\,\alpha_r\,;\,\mrz \\
\beta_1\,\dots\,\beta_s
\end{array}
\right] =
\sum_{n=0}^{\infty}\,
\frac{(\alpha_1\,;\,q)_n\,\dots\,(\alpha_r\,;\,q)_n}{(\beta_1\,;\,q)_n\,\dots\,(\beta_{s+1}\,;\,q)_n}\,\Bigl[ (-1)^n\,q^{n\,(n-1)/2} \Bigr]^{1 + s - r}\mrz^n \spc
\]
whit $\beta_{s+1} = q$ and the $q\,$-shifted factorial defined by
\bq
(\alpha\,;,q)_n = \prod_{k=0}^{n-1}\,\lpar 1 - \alpha\,q^k \rpar \spp
\eq
The basic hypergeometric series was first introduced by Heine and was later generalized by
Ramanujan. For the case of interest we will use the shorthand notation
\[
\upphi\lpar a\,,\,b\,;\,c\,;\,q\,,\,\mrz \rpar =
{}_{2}\upphi_{1} \left[
\begin{array}{c}
a\,,\,b\,;\,\mrz \\
c
\end{array}
\right]  \spp
\]
Furthermore, we introduce
\bq
\upPhi\lpar x\,,\,y\,;\,q \rpar =
\upphi\lpar q\,,\,y\,q\,;\,y\,q^2\,;\,q\,,\,x\,q \rpar \spp
\label{bphi}
\eq
If $| q | < 1$ the $\upphi$ series converges absolutely for $| z | < 1$. The series also converges
absolutely if $| q | > 1$ and $| \mrz | < | c\,q |/| a\,b |$. 
With $a = q, b = y\,q, c = y\,q^2$ and $\mrz = x\,q$ we have that the basic hypergeometric series
converges absolutely if
\bei

\item $| q | < 1$ and $| x\,q | < 1$,

\item $| q | > 1$ and $| x | < 1$, which is a special case of $| \mrz | < | c\,q |/| a\,b |$.

\eei 
In the following we will always assume that $| q | < 1$. Indeed, using $p = 1/q$ and
\bq
(a\,;\,q)_n = (a^{-1}\,;\,p)_n\,( - a )^n\,p^{n\,(n-1)/2} \spc
\eq
one obtains the following relation,
\bq
\upphi\lpar a\,,\,b\,;\,c\,;\,q\,,\,\mrz \rpar =
\upphi\lpar \frac{1}{a}\,,\,\frac{1}{b}\,;\,\frac{1}{c}\,;\,p\,,\,
\frac{a\,b}{c}\,p\,\mrz \rpar \spp
\label{inv}
\eq
Using \eqn{inv} we obtain
\bq
\ELi{0}{0}{x}{y}{q} = \frac{x\,y}{p - y}\,
\upphi\lpar p\,,\,\frac{p}{y}\,;\,\frac{p^2}{y}\,;\,p\,,\,x \rpar \spc
\quad p = \frac{1}{q} \spc
\eq
and the inversion formula
\bq
\ELi{0}{0}{\frac{1}{x}}{\frac{1}{y}}{q} = \frac{1}{x\,(y\,p - 1)}\,\Bphi{\frac{1}{x\,p}}{y}{p} =
- \frac{1}{y}\,\ELi{0}{0}{\frac{1}{x\,p}}{y}{p} \spc
\quad p = \frac{1}{q} \spp
\eq
The next problem is the continuation of the $\upphi$ series into the complex $\mrz$ plane and
the extension to complex $q$ inside the unit disc. We have two alternatives, analytic
continuation and recursion relations. First we present an auxiliary relation that will be 
useful in the continuation of $\upphi$ 
\subsection{q-contiguous relations}
There are several $q\,$-contiguous relations for $\upphi$ and one will be used extensively in 
the rest of this work, see Eq.~(A.10) of \Bref{IL} (see also $1.10$ of \Bref{GR}~\footnote{This 
relation was introduced by E.~Heine~\cite{Heine} in $1847$.}); using a shorthand notation, \ie
\bq
\upphi = {}_2\upphi_1\lpar a\,,\,b\,;\,c\,;\,q\,,\,\mrz \rpar
\qquad
\upphi\lpar c_{\pm n} \rpar = {}_2\upphi_1\lpar a\,,\,b\,;\,c\,q^{\pm n}\,;\,q\,,\,\mrz \rpar \spc
\eq
one derives
\bqa
q\,(1 - c\,q)\,(1 - c)\,(c\,q - a\,b\,\mrz)\,\upphi &=&
(1 - c\,q)\,\Bigl\{ (q^2\,c\,(1 - c) + 
        \Bigl[ (c\,q\,(b + a) - (1 + q)\,a\,b \Bigr]\,\mrz\Bigr\}\,\upphi\lpar c_{+1} \rpar 
\nl
{}&+& (c\,q - a)\,(c\,q - b)\,\mrz\,\upphi\lpar c_{+2} \rpar \spp
\label{qcr}
\eqa
\subsection{Analytic continuation}
A detailed discussion of the analytic continuation of basic hypergeometric series 
can be found in \Bref{GR} (Sects.~4.2-4.10) and in \Bref{Ruij} (Theorem~4.1).
From \Bref{GR} we can use the following analytic continuation
\bqa
{}_2\upphi_{1}\lpar a\,,\,b\,;\,c\,;\,q\,,\,\mrz \rpar &=&
\frac{(b\,,\,c/a\,;\,q)_{\infty}\;
      (a \mrz\,,\,q/a \mrz\,;\,q)_{\infty}}
     {(c\,,\,b/a\,;\,q)_{\infty}\;
      (\mrz\,,\,q/\mrz\,;\,q)_{\infty}}\;
     {}_2\upphi_{1}\lpar a\,,\,a\,q/c\,;\,a\,q/b\,;\,q\,,\,c\,q/a\,b\,\mrz \rpar 
\nl
{}&+&
\frac{(a\,,\,c/b\,;\,q)_{\infty}\;
      (b \mrz\,,\,q/b \mrz\,;\,q)_{\infty}}
     {(c\,,\,a/b\,;\,q)_{\infty}\;
      (\mrz\,,\,q/\mrz\,;\,q)_{\infty}}\;
     {}_2\upphi_{1}\lpar b\,,\,b\,q/c\,;\,b\,q/a\,;\,q\,,\,c\,q/a\,b\,\mrz \rpar \spc
\label{acont}
\eqa
where $| \mathrm{arg}( - \mrz ) | < \pi$, $c$ and $a/b$ are not integer powers of $q$ and
$a, b, \mrz \not= 0$. Furthermore,
\bq
\lpar a_1\,,\,a_2\,;\,q \rpar_{\infty} = 
\lpar a_1\,;\,q \rpar_{\infty}\,\lpar a_2\,;\,q \rpar_{\infty} \spc
\qquad
\lpar a\,;\,q \rpar_{\infty} = \prod_{k=0}^{\infty}\,\lpar 1 - a\,q^k \rpar \spp
\eq
It is worth noting that the coefficients in \eqn{acont} are reducible to ${}_1\upphi_0$ 
functions, \eg
\bq
\frac{(c/a\,;\,q)_{\infty}}{(c\,;\,q)_{\infty}} = \prod_{k=0}^{\infty}\,
\frac{1 - c/a\,q^k}{1 - c\,q^k} = {}_1\upphi_0\lpar \frac{1}{a}\,;\,q\,,\,c \rpar \spc
\eq
\etc A more convenient way to compute the $q\,$-shifted factorial is given by
\bq
(z\,:\,q)_{\infty} = \sum_{n=0}^{\infty}\,\frac{q^{n\,{n-1}/2}}{(q\,;\,q)_n}\,
\lpar - z \rpar^n \spc \qquad
| z | < \infty \spp
\eq
The condition $| \mathrm{arg}( - \mrz ) | < \pi$ must be understood as follows:
The Mellin transform of $\upphi$ (for $\mrz \in \Rf, z > 0$) is a Cauchy Principal Value.
With $| q | < 1$ the two series in \eqn{acont} can be used for $| \mrz | > \mrz_0$ with 
$\mrz_0 = | c/(a\,b)\,q |$. If $| \mrz | < \mrz_0$ we can use \eqn{qcr}; repeated applications 
transform $c$ into $| q^n\,c |$ until a value of $n$ is reached for which 
$\mrz_n = | q^{n + 1}\,c/(a\,b) |$ is such that $| \mrz | > \mrz_n$.

In Sect.~(4.8) of \Bref{GR} it is shown that extension to complex $q$ inside the unit disk
is possible, provided that
\bq
| \mbox{arg}( - \mrz ) - \frac{\omega_2}{\omega_1}\,\ln | \mrz |\,| < \pi \spc
\eq
where $\ln q = - \omega_1 - i\,\omega2$. As seen in the complex $mrz$ plane the condition
is represented by a spiral of equation 
\bq
r = \exp\{\frac{\omega_1}{\omega_2}\,\theta \} \spc \qquad
\mrz= r\,e^{\theta} \spp
\label{spiral}
\eq
It remains to study the convergence for $| q | = 1$. In \Bref{Oshima} a condition is
presented so that the radius of convergence is positive; furthermore, the numbers $q$ with
positive radius are densely distributed on the unit circle.
Obviously, in our case ($a = q$, $b= y\,q$ and $c = y\,q^2$), $q = 1$ reduces any 
$\mathrm{ELi}$ function to a product of polylogarithms.
\subsection{Basic hypergeometric equation}
In the context of functional equations the basic hypergeometric series provides a solution
to a second order $q\,$-difference equation, called the basic hypergeometric equation,
see \Brefs{Roques,Ruij}. 

We will now show how $\upphi$ can be computed to arbitrary precision, using a theorem
proved in \Bref{CHM}.
\begin{theorem}[Chen, Hou, Mu]
Let $f(\mrz)$ be a continuous function defined for $| \mrz | < r$ and $d \ge 2$ an integer.
Suppose that 
\bq
f(\mrz) = \sum_{n=1}^{d}\,a_n(\mrz)\,f(\mrz\,q^n) \spc
\eq
with $a_n(0) = \mrw_n$. Suppose that exists a real number $\mrM > 0$ such that
\bq
| a_n(\mrz) - \mrw_n | \le \mrM\,| \mrz |, \quad 1 \le n \le d \spc
\eq
and
\bq
| \mrw_d | + | \mrw_{d-1} + \mrw_d | +\,\dots\, | \mrw_2 +\,\dots\,\mrw_d | < 1 \spc
\qquad
\mrw_1 +\,\dots\,+ \mrw_d = 1 \spp
\label{ccond}
\eq
The $f(\mrz)$ is uniquely determined by $f(0)$ and the functions $a_n(\mrz)$.
\end{theorem}
With
\bq
\mrF(\mrz) = {}_2\upphi_{1}\lpar a\,,\,b\,;\,c\,;\,q\,,\,\mrz \rpar 
\eq
we have
\bq
\mrF(\mrz) = \frac{(a + b)\,q\,\mrz - c - q}{q\,(\mrz - 1)}\,\mrF(\mrz\,q) +
\frac{c - q\,a\,b\,\mrz}{q\,(\mrz - 1)}\,\mrF(\mrz\,q^2) \spc
\label{bhrr}
\eq
thus, by the theorem and for $| q | < 1$ and $| c/q | < 1$ (from \eqn{ccond}) we can determine 
uniquely $\mrF(\mrz)$ by $\mrF(0)$ and the $q\,$-difference equation (\eqn{bhrr}), \ie we define
\[
\left\{
\begin{array}{l}
\mrA^{(i)}_{n+1} = a_i(\mrz\,q^{n+1})\,\mrA^{(1)}_n + \mrA^{(i+1)}_n \spc \quad
1 \le i < 2 \\
\mrA^{(2)}_{n+1} = a_2(\mrz\,q^{n+1})\,\mrA^{(1)}_n
\end{array}
\right.
\]
with $\mrA^{(i)}_0 = a_i(\mrz)$ and obtain
\bq
\mrF(\mrz) = \mrF(0)\,\sum_{i=1}^{2}\,\lim_{n \to \infty}\,\mrA^{(i)}_n \spp
\eq
High accuracy can be obtained by computing
\bq
\mrF_{\mrN}(\mrz) = \mrF(0)\,\sum_{i=1}^{2}\,\mrA^{(i)}_{\mrN} \spc
\eq
for $\mrN$ high enough.
Application to ${}_2\upphi_{1}\lpar q\,,\,y\,q\,;\,y\,q^2\,;\,q\,,\,x\,q \rpar$ requires
$| y\,q | < 1$. 

If $| q | > 1$ we can use the $q\,$-difference equation downword. In this case
\bq
\mrF(\mrz) = \frac{(c + q)\,q - (a + b)\,\mrz}{c\,q - a\,b\,\mrz}\,\mrF(q^{-1}\,\mrz) +
\frac{\mrz - q^2}{c\,q - a\,b\,\mrz}\,\mrF(q^{-2}\,\mrz) \spc
\eq
and
\[
\left\{
\begin{array}{l}
\mrA^{(i)}_{n+1} = a_i(\mrz\,q^{-n-1})\,\mrA^{(1)}_n + \mrA^{(i+1)}_n \spc \quad
1 \le i < 2 \\
\mrA^{(2)}_{n+1} = a_2(\mrz\,q^{-n-1})\,\mrA^{(1)}_n
\end{array}
\right.
\]
which requires $| q/c | < 1$. In both cases the $q\,$-difference equation determines $\mrF$ without
limitations in the $\mrz$ complex plane. 

To summarize, we can compute $\upphi$ by using the sum of the series inside the
circle of convergence (and their analytic continuation) or by using the $q\,$-difference equation.
The advantage in using the latter is no limitation on $\mrz$ but
\bq
| q | < 1\;\;\mathrm{and}\;\; | c/q | < 1
\quad \mathrm{or} \quad
| q | > 1\;\;\mathrm{and}\;\; | q/c | < 1 \spp
\label{rrc}
\eq
To be precise, the function defined by \eqns{bhrr}{rrc} is a meromorphic continuation
of $\upphi$ with simple poles located at $\mrz = q^{- n}, \;\; n \in \Zf^*$. 
Seen as a function of $\alpha= \ln(z)/(2\,i\,\pi)$ the continuation shows poles
at $2\,i\,\pi\,\alpha = n\,\omega$ where $\omega= - \ln q$.
What to do when we are outside the two regions of applicability? Instead of using analytic
continuation we can do the following: assume that $| q | < 1$ but $| c/q | > 1$.
We can use \eqn{qcr}; repeated applications transform $| c/q |$ into $| q^n\,c |$ until a value 
of $n$ is reached for which $| q^n\,c | < 1$. 
Similar situation when $| q | > 1$ and $| q/c | > 1$ where repeated applications transform 
$| q/c |$ into $| q^{-n}\,c |$ until a value of $n$ is reached for which $| q^{-n}\,c | < 1$. 

It is worth noting that the $\upPhi$ function satisfies
\bq
\Bphi{x}{y}{q} = \Bigl[ \frac{1 - q^2\,x}{1 - q\,x} + q\,y \Bigr]\,\Bphi{x\,q}{y}{q} -
                 q\,y\frac{1 - q^2\,x}{1 - q\,x}\,\Bphi{x\,q^2}{y}{q} = 
x\,q\,\frac{1 - q\,y}{1 - q^2\,y}\,\Bphi{x}{y\,q}{q} + 1  \spc
\eq
which are the $q\,$-shift along $x$ and $y$.
\subsection{Poles}
Using \eqn{eli00def} and \eqn{bhrr} we conclude that $\ELi{0}{0}{x}{y}{q}$ has simple poles 
located at 
\bq
x\,q = q^{- n} \spc \quad
y\,q = q^{- n} \spc \qquad
n \in \Zf^* \spp 
\eq
Poles in the complex $y\,$-plane can also be analyzed by using \eqn{eli00def} and
\eqn{qcr}. Indeed, from \eqn{qcr} we obtain
\bqa
\Bphi{x}{y}{q} &=&
       \frac{1}{1 - x}\,\Bigl[
          \frac{1 + y}{1 - q^2\,y}\,x\,q -
          \frac{1 + q}{q}\,x
          + 1
          \Bigr]\,
          \upphi\lpar q\,,\,y\,q\,;\,y\,q^3\,;\,q\,,\,x\,q \rpar
\nl
{}&+& \frac{x}{1 - x}\,\frac{1 - q^2}{1 - q^3\,y}\,\frac{1}{q}\,
          \upphi\lpar q\,,\,y\,q\,;\,y\,q^4\,;\,q\,,\,x\,q \rpar \spp
\eqa
Using $\Bphi{x}{\frac{1}{q}}{q} = 1$ we have a simple pole of $\ELi{0}{0}{x}{y}{q}$ at 
$y = 1/q$ with residue $- x/q$ (the pole at $x = 1/q$ has residue $- y/q$). Repeated applications 
of the $q\,$-contiguous relation exhibit the poles at $y\,q = q^{- n}, n \in \Zf^*$. 

Residues can be computed according to the following chain
\bq
\Bphi{x}{y}{q} = \frac{\mrR_1}{x - \frac{1}{q}} + \mathrm{Reg}_1\lpar x\,,\,y\,;\,q \rpar = 
\frac{\mrR_2}{x - \frac{1}{q^2}} + \mathrm{Reg}_2\lpar x\,,\,y\,;\,q \rpar = \mbox{\etc} \spc
\eq
where the ``regular'' part admits a Taylor expansion around $x = 1/q$, $x = 1/q^2$ etc.
\bqa
\mrR_1 &=& \lpar 1 - \frac{1}{q} \rpar\,\Bphi{1}{y}{q} +
            y\,\lpar 1 - q \rpar\,\Bphi{q}{y}{q} \spc
\nl 
\mrR_2 &=& y\,\lpar 1 - \frac{1}{q} \rpar\,\Bphi{1}{y}{q} +
            y^2\,\lpar 1 - q \rpar\,\Bphi{q}{y}{q} \spc
\nl
{}&=& \mbox{\etc}
\eqa
Using the series
\bq
\Bphi{q^k}{y}{q} = (1 - y\,q)\,\sum_{n=0}^{\infty}\,\frac{q^{n\,(k + 1)}}{1 - y\,q^{n+1}} \spc
\eq
we obtain
\bq
\mrR_1= (y\,q - 1)\,\frac{1}{q} \spc \qquad
\mrR_1= (y\,q - 1)\,\frac{y}{q} \spc \qquad
\mbox{\etc}
\eq
showing the following residues for $\ELi{0}{0}{x}{y}{q}$:
\bq
\mathrm{Res}_{x = 1/q^n}\,\ELi{0}{0}{x}{y}{q} = - \frac{y}{q} \;,\; 
                                                - \frac{y^2}{q^2} \;,\;
                                                - \frac{y^3}{q^3} \;,\;
                                                  \dots
\qquad \mbox{for} \quad n = 1\,,\,2\,,\,3\,,\,\dots
\eq
The isolation of simple poles in $\ELi{0}{0}{x}{y}{q}$ is crucial in order to compute
Elliptic polylogarithms of higher depth.
\section{Elliptic polylogarithms of higher depth}
In this Section we show how to compute arbitrary elliptic polylogarithms, in particular how to
identify their branch points (their multi-valued component).

The procedure is facilitated by the fact that both the basic hypergeometric equation and the
$q\,$-contiguous relation allows to isolate the (simple) poles of 
$\ELi{0}{0}{x}{y}{q}$ with a remainder given by a ``+'' distribution. 

Introducing the usual $i\,\ep$ prescription we obtain a general recipe for computing elliptic 
polylogarithms ``on the cuts'' ($x$ and or $y$ real and greater than $1/q$). In the following 
we discuss few explicit examples.

{$\underline{\bf{\ELi{1}{0}{x}{y}{q}}}$}

From \eqn{reda} we obtain
\bq
\ELi{1}{0}{x}{y}{q} = \frac{x y q}{1 - y\,q}\,\int_0^1\,d\mrz\,
\Bphi{\mrz\,x}{y}{q} \spp
\eq
For $x\,q \in \Rf$ the integral is defined when $x < 1/q$, otherwise it is understood that
$x \to x \pm i\,\ep$ where $\ep \to 0_{+}$. Consider the case
\bq
\frac{1}{q} < x < \frac{1}{q^2} \spp
\label{cond}
\eq
From \eqn{bhrr} we derive
\bq
\Bphi{\mrz\,x}{y}{q} = \sum_{i=1,2}\,\frac{\mrt_i(\mrz)}{\mrz\,x\,q - 1}\,
\Bphi{\mrz\,x\,\,q^i}{y}{q} \spc
\label{polex}
\eq
From \eqn{cond} it follows that no pole of the two $\upphi$ functions in \eqn{polex}
appears for $z \in [0,1]$; therefore the two $\upphi$ in the r.h.s of \eqn{polex}
can be evaluated according to the strategy outlined in the previous Sections. 
Using
\bq
t_1\lpar \frac{1}{x\,q} \rpar = q - 1 \spc \qquad
t_2\lpar \frac{1}{x\,q} \rpar = y\,q\,\lpar 1 - q \rpar \spc
\eq
we can write
\bq
\int_0^1\,d\mrz\,\Bphi{\mrz\,x}{y}{q} =
\frac{q - 1}{x\,q}\,\ln\lpar 1 - x\,q \rpar\,
\Bigl[ \Bphi{1}{y}{q} - y\,q\,\Bphi{q}{y}{q} \Bigr] + \mrS \spc
\label{logb}
\eq
where the ``subtraction'' term is
\bq
\mrS = \int_0^1\,\frac{d\mrz}{\mrz\,x\,q - 1}\,
\sum_{i=1,2}\,\Bigl[ \mrt_i(\mrz)\,\Bphi{\mrz\,x\,q^i}{y}{q} 
- \mrt_i(\frac{1}{x\,q})\,\Bphi{q^{i-1}}{y}{q} \Bigr] \spc
\label{sub}
\eq
and $x = x \pm i\,\ep$.  Note that
\bq
\Bphi{1}{y}{q} - y\,q\,\Bphi{q}{y}{q} = \frac{1 - y\,q}{1 - q} \spp
\eq
If $q^{-2} < x < q^{-3}$ we can iterate once more obtaining an
additional $\ln(1 - x\,q^2)$ \etc
The function $\mrS$ defined in \eqn{sub} is a ``+'' distribution which has simple poles in 
the $y\,$-plane.
The explicit result is as follows:
\bq
\ELi{1}{0}{x}{y}{q} = \ELic{1}{1}{0}{x}{y}{q} + \ELir{1}{1}{0}{x}{y}{q} \spc
\qquad \frac{1}{q} < x < \frac{1}{q^2} \spc
\eq
where the ``cut'' part is
\bq
\ELic{1}{1}{0}{x}{y}{q} = - y\,\ln(1 - x\,q) \spc
\eq
while the ``restr'' is
\bqa
\ELir{1}{1}{0}{x}{y}{q} &=& \frac{x\,y\,q}{1 - y\,q}\,
\int_0^1\,d\mru\,\Bigl[
\frac{\mrF_1(u) - 1 + y\,q}{1 - x\,q\,u} + y\,q\,\Bphi{x\,q\,u}{y}{q} \Bigr] \spc
\nl
\mrF_1(u) &=& \lpar 1 - x\,q^2\,u \rpar\,\Bigl[ \Bphi{x\,q\,u}{y}{q} -
y\,q\,\Bphi{x\,q^2\,u}{y}{q} \Bigr] \spp
\eqa
The second iteration gives
\bq
\ELi{1}{0}{x}{y}{q} = \ELic{2}{1}{0}{x}{y}{q} + \ELir{2}{1}{0}{x}{y}{q} \spc
\qquad \frac{1}{q^2} < x < \frac{1}{q^3} \spc
\eq
\bq
\ELic{2}{1}{0}{x}{y}{q} =  - y\,\ln(1 - x\,q) - y^2\,\ln(1 - x\,q^2) \spc
\eq
\bqa
\ELir{2}{1}{0}{x}{y}{q} &=& \frac{x\,y\,q}{1 - y\,q}\,
\int_0^1\,d\mru\,\Bigl[
\frac{\mrF_{21}(\mru) - \mrF_{21}(\frac{1}{x q})}{x\,\mru - \frac{1}{q}} +
\frac{\mrF_{22}(\mru) - \mrF_{22}(\frac{1}{x q^2})}{x\,\mru - \frac{1}{q^2}} +
\mrR_2(\mru) \Bigr] \spc
\eqa
\bqa
\mrF_{21}(u) &=&
 \Bigl[ 1 + q\,\lpar 1 + y \rpar \Bigr]\,\frac{q - 1}{q}\,\Bphi{x\,q^2}{y}{q}
 - y\,\Bigl\{\Bigl[ 1 + q\,\lpar q + 1 \rpar\,\lpar 1 + y \rpar \Bigr]\,\frac{q - 1}{q + 1} + 
          \lpar q^2 - 1 \rpar\Bigr\}\,\Bphi{x\,q^3}{y}{q}
\nl
{}&+& \lpar 1 + q + q^2 \rpar\,q\,y^2\,\frac{q - 1}{q + 1}\,\Bphi{x\,q^4}{y}{q} \spc
\nl\nl
\mrF_{22}(u) &=&  y\,\lpar q - 1\rpar\,
\Bigl[ \frac{1}{q}\,\Bphi{x\,q^2}{y}{q} - y\,\Bphi{x\,q^3}{y}{q} \Bigr] \spc
\nl\nl
\mrR_2(u) &=&
  + q^2\,\lpar 1 + y \rpar^2\,\Bphi{x\,q^2}{y}{q}
  - y\,\Bigl[ \frac{1}{q}\,\frac{q - 1}{q + 1}\,\frac{1}{x\,u - \frac{1}{q^3}} + 
       2\,q^3\,\lpar 1 + y \rpar \Bigr]\,\Bphi{x\,q^3}{y}{q} 
\nl
{}&+& y^2\,\lpar \frac{q - 1}{q + 1}\,
      \frac{1}{x\,u - \frac{1}{q^3}} + q^4 \rpar\,\Bphi{x\,q^4}{y}{q} \spp
\eqa
We continue with other examples.

{$\underline{\bf{\ELi{0}{1}{x}{y}{q}}}$}

Similar results follow from \eqn{eli00def}, \ie by using the $x\,,\,y$ symmetry of
$\ELi{0}{0}{x}{y}{q}$.

{$\underline{\bf{\ELi{2}{0}{x}{y}{q}}}$}

We can use the same derivation as before obtaining a result similar to the one in \eqn{logb}
where we replace
\bq
\ln\lpar 1 - x\,q \rpar \to -\,\mathrm{Li}_2 \lpar x\,q \rpar \spc
\eq
where $\li{2}{z}$ is the dilogarithm; for $\ELi{n}{0}{x}{y}{q}$ with $n > 2$ we obtain
polylogarithms. The explicit result is as follows:
\bq
\ELi{2}{0}{x}{y}{q} = \ELic{1}{2}{0}{x}{y}{q} + \ELir{1}{2}{0}{x}{y}{q} \spc
\qquad \frac{1}{q} < x < \frac{1}{q^2} \spc
\eq
where the ``cut'' part is
\bq
\ELic{1}{2}{0}{x}{y}{q} = y\,\li{2}{x\,q} \spc
\eq
while the ``rest'' part is
\bqa
\ELir{1}{2}{0}{x}{y}{q} &= & \frac{x\,y\,q}{1 - y\,q}\,\int_0^1\,\frac{d\mrv}{\mrv}\,
\int_0^{\mrv}\,d\mru\,\Bigl[
\frac{\mrF_1(u) - 1 + y\,q}{1 - x\,q\,u} + y\,q\,\Bphi{x\,q\,u}{y}{q} \Bigr] \spc
\nl
\mrF_1(u) &=& \lpar 1 - x\,q^2\,u \rpar\,\Bigl[ \Bphi{x\,q\,u}{y}{q} -
y\,q\,\Bphi{x\,q^2\,u}{y}{q} \Bigr] \spc
\eqa
When both $n$ and $m$ are different from zero the derivation requires isolating 
$x\,$-poles and $y\,$-poles.

{$\underline{\bf{\ELi{1}{1}{x}{y}{q}}}$}

This case requires additional work. Consider the integral
\bq
\int_0^1\,d\mrz_1 d\mrz_2\,\frac{1}{1 - \mrz_2\,y\,q}\,
\Bphi{\mrz_1\,x}{\mrz_2\,y\,q}{q} \spp
\label{oneone}
\eq
When $y\,,\,q \in \Rf$ the integral is defined for $y < 1/q$, otherwise it is understood
that $y \to y \pm o\,\ep$ with $\ep \to 0_{+}$.
By using \eqn{qcr} we obtain
\bqa
\upphi(c) &=&  \Bigl[  q\,(1 - c\,q)\,(1 - c)\,(c\,q - a\,b\,u) \Bigr]^{-1}\,\Bigl\{
              (1 - c\,q)\,\Bigl[
              + (b + a)\,c\,q\,u
              - (1 + q)\,a\,b\,u
              + (1 - c)\,c\,q^2
              \Bigr]\,\upphi(c\,q) 
\nl
{}&+& (c\,q - a)\,(c\,q - b)\,u\,\upphi(c\,q^2) \Bigr\} \spc
\eqa
where $\upphi(c) = \upphi\lpar a\,,\,b\,;\,c\,;\,q\,,\,u \rpar$.
Next we replace $a = q$, $b = \mrz_2\,y\,q$, $c = \mrz_2\,y\,q^2$ and $u = \mrz_1\,x\,q$. If 
$1/q < y < 1/q^2$ we replace $\mrz_2 = 1/(y\,q)$ in the $\upphi$ function of \eqn{oneone} and 
obtain a logarithmic part
\bq
- \,\frac{1}{y\,q}\,\ln\lpar 1 - y\,q \rpar \spc
\eq
as well as a ``subtraction'' part. Note that $b = 1$ gives a terminating series, \ie
\bq
\upphi\lpar q\,,\,1\,;\,q^3\,;\,q\,,\,\mrz_1\,x\,q \rpar = 1 \spp
\eq

{$\underline{\bf{\mathrm{ELi}_{0\,,\,0\,;\,0\,,\,0\,;\,2}}}$} 
This function is defined through
\bq
\mathrm{ELi}_{0\,,\,0\,;\,0\,,\,0\,;\,2} 
\lpar x_1\,,\,x_2\,;\,y_1\,,\,y_2\,;\,q \rpar =
\int_0^1 \frac{d\mrz}{\mrz}\,
\ELi{0}{0}{x_1}{y_1}{\mrz\,q}\,
\ELi{0}{0}{x_2}{y_2}{\mrz\,q} \spp
\eq
The integrand has poles in the complex $\mrz\,$-plane located at
\[
\mrz = \left\{
\begin{array}{l}
\frac{1}{x_i\,q^{n+1}} \\
                       \\
\frac{1}{y_i\,q^{m+1}} 
\end{array}
\qquad i= 1,2 \quad n, m \in \Zf^*
\right.
\]
Poles of the first series are isolated by using the $q\,$-difference equation while
those in the second series are isolated by using the $q\,$-contiguous relation. All poles
are simple as long as none of the ratios $x_i/x_j, y_i/y_j$ and $x_i/y_j$ is equal to an
integer power of $q$.
\subsection{Mixed hypergeometric series}
When $| q | < 1$ and $x < 1/q$ we can write $\ELi{n}{0}{x}{y}{q}$ (or $y < 1/q$ and 
$\ELi{0}{n}{x}{y}{q}$) as mixed hypergeometric series, \eg
\bq
\ELi{1}{0}{x}{y}{q} = 
{}_{2\,,\,2}\upphi_{1\,,\,1}\lpar q\,,\,y\,q\,;\,1\,,\,1\,;\,y\,q^2\,,\,2\,;\,x\,q \rpar = 
\sum_{n=0}^{\infty}\,\frac{(y\,q\,;\,q)_n\,(1)_n\,(1)_n}{(y\,q^2\,;\,q)_n\,(2)_n}\,
\frac{(x\,q)^n}{n\,!} \spc
\eq
where $(a)_n = \Gamma(a+n)/\Gamma(a)$ is the Pochhammer symbol. For a previous definition of
mixed hypergeometric series see \Bref{Khans}.
\subsection{Barnes contour integrals}
For $0 < q < 1$ we can write
\bq
\ELi{1}{0}{x}{y}{q} = - \frac{x\,y\,q}{2\,\pi\,i}\int_{\Gamma}\,d\mrs
\frac{\pi}{\sin\,\pi s}\,\frac{( - x\,q )^s}{s + 1}\,\frac{1}{1 - y\,q^{s+1}} \spc
\label{MB}
\eq
where $| x\,q | < 1$ and $| \mathrm{arg}( - x\,q) | < \pi$. The contour of integration,
denoted by $\Gamma$, runs from $-\,i\,\infty$ to $+\,i\,\infty$ so that the poles at
$s \in \Zf^*$ lie to the right of the contour and the other poles, at $s \in \Zf^{-}$ and
$s = - 1 + (\ln y + 2\,m\,\pi\,i)/omega$ with $\omega= - \ln q$ and $m \in \Zf$,
lie to the left and the latter are at least some $\ep$ ($\ep \to 0_{+}$) distance away from the 
contour.
The r.h.s. of \eqn{MB} defines an analytic function of $\mrz = x\,q$ in 
$| \mathrm{arg}( - z) | < \pi$. Note that \eqn{MB} can be generalized to define the analytic
continuation of $\ELi{n}{0}{x}{y}{q}$ and can be extended to complex $q$ inside the unit
disc (see \eqn{spiral}).
\subsection{Eisenstein-Kronecker series}
The construction of elliptic multiple polylogarithms in \Bref{BL} is largely based
on the Eisenstein-Kronecker series $\mrF(\xi\,,\,\alpha\,,\,\tau)$ defined in their
Sect.~3.4; with
\bq
\mrz = e(\xi) \spc \quad q = e(\tau) \spc \quad \alpha = e(u) \spc
\eq
where $e(\xi) = \exp\{2\,\pi\,i\,\xi\}$ we obtain the following relation with $\upPhi$,
the basic hypergeometric series of \eqn{bphi}:
\bq
\frac{1}{2\,\pi\,i}\,\mrF(\xi\,,\,\alpha\,,\,\tau) = \frac{\mrz}{\mrz - 1} - 
\frac{u\,\mrz\,q}{1 - \mrz\,q}\,\Bphi{u}{\mrz}{q} + 
\frac{\mrz}{u\,(\mrz - q)}\,\Bphi{\frac{1}{u}}{\mrz}{\frac{1}{q}} \spp
\eq
The function $\mrF$ satisfies
\bq
\mrF(\xi + 1\,,\,\alpha\,,\,\tau) = \mrF(\xi\,,\,\alpha\,,\,\tau) \spc
\qquad 
\mrF(\xi + \tau\,,\,\alpha\,,\,\tau) = \frac{1}{\alpha}\,\mrF(\xi\,,\,\alpha\,,\,\tau) \spc
\eq
\ie quasi-periodicity.
\section{Conclusions}
\begin{figure}[hbt]
 \begin{center}
 \includegraphics[width=1.\textwidth]{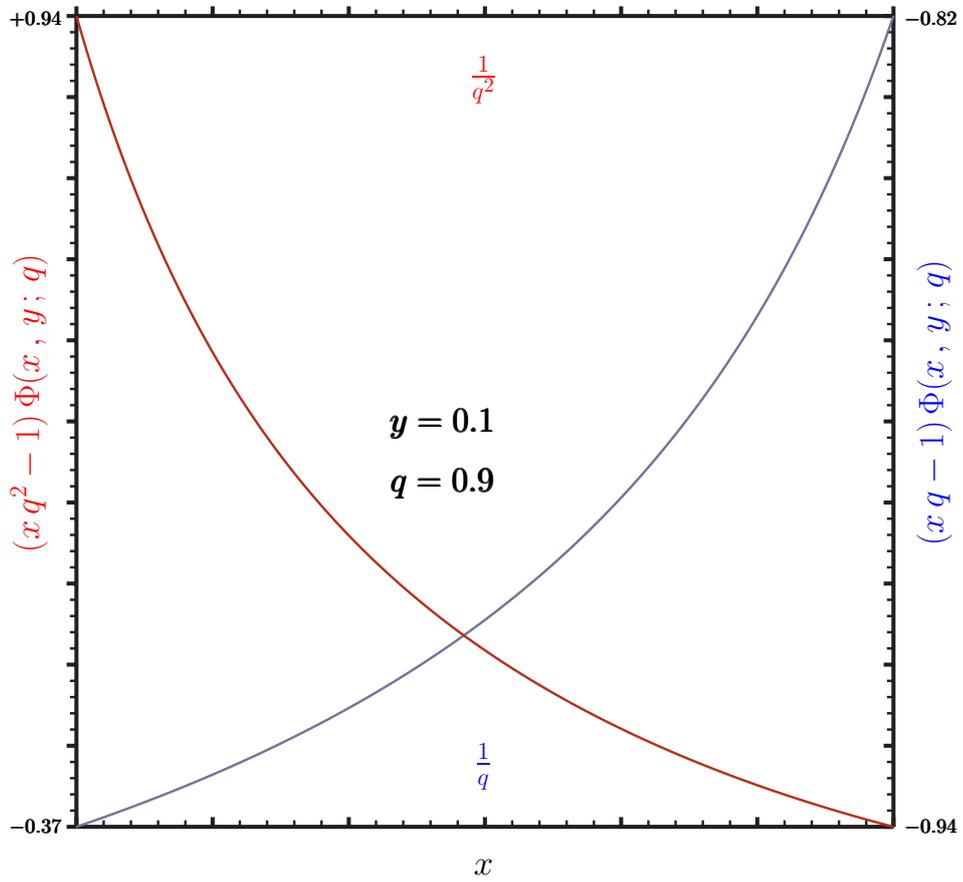}
 \vspace{-7.cm}
 \caption{Behavior of $\Bphi{x}{y}{q}$, defined in \eqn{bphi}, around the poles at $x = 1/q$ 
(blue curve) and $x = 1/q^2$ (red curve).
\label{Fig}
         }
 \end{center}
\end{figure}

\begin{figure}[hbt]
 \vspace{-4.cm}
 \begin{center}
 \includegraphics[width=1.\textwidth]{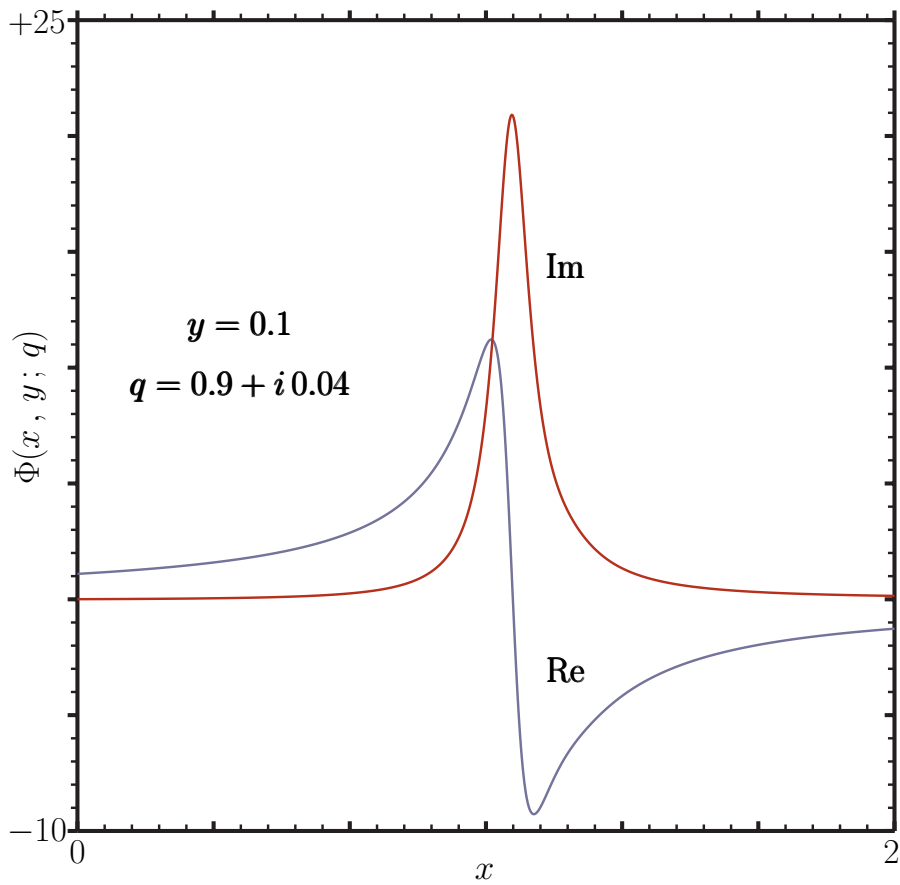}
 \vspace{-7.cm}
 \caption{Behavior of $\Bphi{x}{y}{q}$, defined in \eqn{bphi}, as a function of $x$ for 
$y= 0.1$ and $q= 0.9 + i\,0.04$.
\label{FigC}
         }
 \end{center}
\end{figure}

\begin{figure}[hbt]
 \vspace{-4.cm}
 \begin{center}
 \includegraphics[width=1.\textwidth]{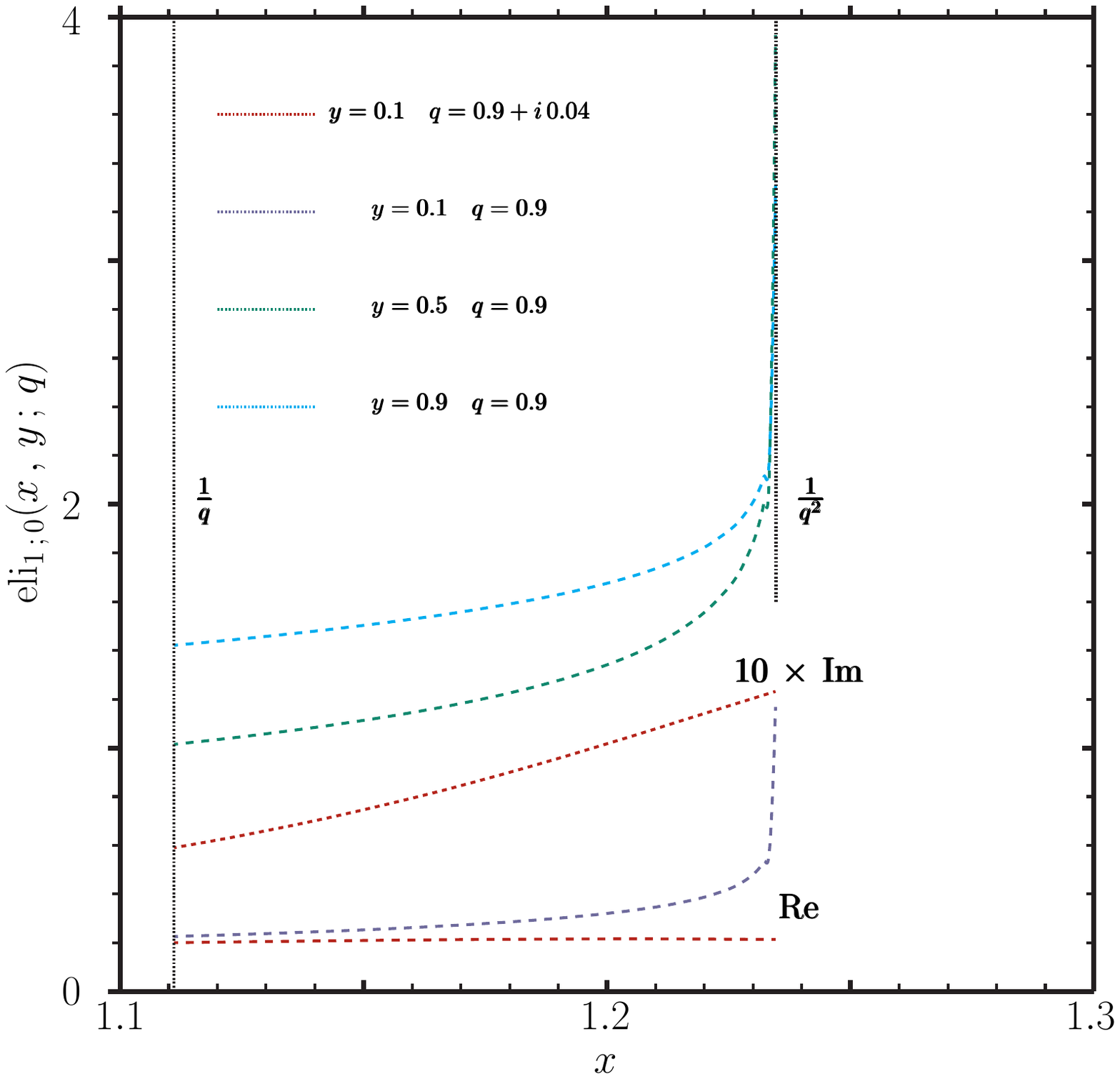}
 \vspace{-7.cm}
 \caption{The function $\mathrm{eli}_{1\,;\,0}(x\,,\,y\,;\,q)$, \eqn{seli}, for different values of
$y$ and $q$ and $1/q_{\mrr} < x < 1/q^2_{\mrr}$, $q_{\mrr} = \Re\,q$.
\label{Figel10}
         }
 \end{center}
\end{figure}

We have established a connection between elliptic polylogarithms and basic hypergeometric
functions, providing a framework for high-precision numerical evaluation of the
$\mathrm{Eli}$ functions. Outside the region of convergence of the series the numerical
evaluation uses analytic continuation via Watson's contour integral representation, basic 
hypergeometric equation and $q\,$-contiguous relations.
As an example we show in Fig.~\ref{Fig} the behavior of $\Bphi{x}{y}{q}$, \eqn{bphi},  around the
poles at $x = 1/q$ (blue curve) and $x= 1/q^2$ (red curve) for $y = 0.1$ and $q= 0.9$.
In Fig.~\ref{FigC} we show the real and the imaginary parts of $\Bphi{x}{y}{q}$ for
$y= 0.1$ and $q = 0.9 + i\,0.04$.
In Fig.\ref{Figel10} we show $\mathrm{eli}_{1\,;\,0}$, defined by
\bq
\ELir{1}{1}{0}{x}{y}{q} =  \frac{x\,y\,q}{1 - y\,q}\,
\mathrm{eli}_{1\,;\,0}\lpar x\,,\,y\,;\,q \rpar \spc
\label{seli}
\eq
for different values of $y$ and $q$ and $1/q_{\mrr} < x < 1/q^2_{\mrr}$, $q_{\mrr} = \Re\,q$.

 \clearpage
\bibliographystyle{elsarticle-num}
\bibliography{EPBHF}

\end{document}